\documentclass[a4paper]{article}
\usepackage{graphicx}
\usepackage[letterpaper, margin=1in]{geometry}
\usepackage{epstopdf, epsfig}
\usepackage{authblk}
\usepackage{color}
\usepackage{amssymb,amsmath}

\title{Extreme waves in crossing sea states}
 
\author[1]{Joseph Brennan}
\author[2]{John M. Dudley}
\author[3]{Fr\'ed\'eric Dias\thanks{Corresponding author, e-mail: frederic.dias@ucd.ie}}

\affil[1,2]{School of Mathematics and Statistics, University College Dublin, Belfield Dublin 4, Ireland}
\affil[3]{Institut FEMTO-ST CNRS, Universit\'e de Franche-Comt\'e, UMR 6174, France}

\date{}%Do not print date

\begin{document}
\maketitle

\begin{abstract}

The evolution of crossing sea states and the emergence of rogue waves in such systems are studied via numerical simulations performed using a higher order spectral method to solve the free surface Euler equations with a flat bottom. Two classes of crossing sea states are analysed: one using directional spectra from the Draupner wave crossing at different angles, another considering a Draupner-like spectra crossed with a narrowband JONSWAP state to model spectral growth between wind sea and swell. These two classes of crossing sea states are constructed using the spectral output of a WAVEWATCH III hindcast on the Draupner rogue wave event. We measure ensemble statistical moments as functions of time, finding that although the crossing angle influences the statistical evolution to some degree, there are no significant third order effects present. Additionally, we pay particular attention to the mean sea level measured beneath extreme crest heights, the elevation of which (set up or set down) is shown to be related to the spectral content in the low wavenumber region of the corresponding spectrum. 
\end{abstract}

\section{Introduction}

Over the last two decades, the rogue wave phenomenon has received acute attention, with extensive research being performed theoretically, numerically and experimentally. One particular area of interest within this field is that of the emergence of rogue waves within crossing sea states, or in other words a sea state composed of two distinct wave systems, propagating with different directions relative to each other. Such a wave system is not an uncommon occurrence in the ocean, and there are notable rogue wave incidents recorded in such states, such as the Suwa Maru (see Tamura \textit{et al.} (2009)), Louis Majesty (see Cavaleri \textit{et al.} (2012)) and Prestige incidents (see Trulsen \textit{et al.} (2015)). The Draupner incident, which is perhaps the most famous rogue wave event, has also been linked (for example, see Adcock \textit{et al.} (2011)) with potential crossing sea activity. More recently, it was conjectured by Fedele \textit{et al.} (2017) that the El Faro rogue wave also occurred in a multidirectional sea state.

We note here that the term ``bimodality" has sometimes been used to describe crossing seas (for example, see Toffoli \textit{et al.} (2006)). However, the classical definition of bimodality refers to a different phenomenon, namely the propagation of more wave energy at an angle to the wind than in the wind direction. Time-resolved measurements of ocean waves have shown a prevalence of directional bimodality at frequencies above twice the peak frequency (for example, see Young \textit{et al.} (1995) and Wang and Hwang \textit{et al.} (2001)). These were confirmed by airborne remote sensing techniques (for example, see Romero and Melville (2010)), and by stereo-video systems (for example, Peureux \textit{et al.} (2018)). The bimodality is apparently caused by the nonlinear cascade of free wave energy from dominant to high frequencies. Bimodality is also found by solving the free-surface Euler equations for the temporal evolution of initially unimodal directional wave spectra (Toffoli \textit{et al.} (2010)). We will not consider bimodality any longer in this paper. 

Theoretically, coupled nonlinear Schr\"{o}dinger (CNLS) equations have been used in the study of crossing wave-trains and their nonlinear interactions, and in particular, the Benjamin--Feir (or modulational) instability. Narrowband wave trains are well known to be susceptible to this, and the instability has been studied thoroughly in terms of the nonlinear Schr\"{o}dinger Equation (NLS) for deep water narrowband wave envelopes. Indeed, the Benjamin--Feir instability has been put forward as a possible generating mechanism for rogue waves, given the appropriate conditions. Onorato \textit{et al.} (2006) derived a set of CNLS equations in 2 + 1 dimensions to study the instability in non-colinearly propagating wave trains, and performed a stability analysis for perturbations confined to the $x-$axis (i.e. propagation in one dimension). Shukla \textit{et al.} (2006), using these equations, performed a similar analysis incorporating perturbations in two directions. Laine-Pearson (2010) developed a theory for the long-wave instability of short-crested waves. Short-crested waves are the resonant interaction of two wave systems each with a different direction of propagation. The stability of these wave interactions is closely associated with the stability of the oblique nonresonant interaction between two waves. By considering the long-wave instability of such waves, Laine-Pearson demonstrated that instability growth rates of two crossing waves can be larger than those given by short-crested waves, concluding that only considering true resonant interactions can underestimate the contribution from unstable crossing sea states to the possible formation of rogue waves. Ruban (2009,2010) considered the formation of rogue waves in crossing seas using a system of equations with weak three-dimensional effects. 

A stability analysis for the CNLS system yields an expression for the instability growth rate in the crossing wave trains, which is in fact dependent on the angle between the directions of propagation of each wave train. Let the wave vectors of the two propagation directions be given by $(k_x,k_y)$ and $(k_x,-k_y$), where $k_x$ is the wave vector component along the $x-$axis and $\pm k_y$ the wave vector components along the $y-$axis. The crossing angle between them is given by $\Omega = 2\tan^{-1}(k_y/k_x)$. The growth rate becomes negative at critical angle $\Omega_c = 2\tan^{-1}(1/\sqrt{2}) = 70.53^{o}$, and so focusing instability was found for angles $0 < \Omega < 70.53^{o}$. The instabilities become defocusing after this point, and nonlinear interactions were found to strengthen as the crossing angle approached $\Omega_c$. 

Masson (1993) considered the nonlinear coupling by way of resonant interactions between swell and wind wave sea states via the Hasselmann equation. It was found that significant energy transfer may take place, causing the swell system to grow at the expense of the short systems. This transfer depends on the crossing angle, reaching its maximum at approximately $40^{\circ}$. Furthermore, the ratio between the peak frequencies of each system influenced the strength of the coupling, with no significant interaction for ratios less than $0.6$. However, given that ocean waves are typically low frequency, this type of coupling is thus negligible unless the spectral peaks are so close together that their double peaked structure is difficult to identify. 

As touched on earlier, it is reasonable to speculate that at the time of both the Draupner event and the El Faro events, the directional spectra actually resulted from the crossing of two spectra travelling at an angle of each other. This speculation is supported by the fact that a positive mean sea level, or `set up', was measured at the apex of both rogue waves. This is quite puzzling, as a `set down' (i.e., a negative mean sea level) is generally expected to be observed beneath such large crests, and this is probably a consequence of the potential crossing sea feature. In this article, we seek to address the short time evolution of crossing sea states, examining the statistical moments linked to nonlinear wave interactions. Two classes of crossing sea states are analysed: one using directional spectra from the Draupner wave crossing at different angles, another considering a Draupner-like spectra crossed with a narrowband JONSWAP state to model spectral growth between wind sea and swell. These two classes of crossing sea states are constructed using the spectral output of a WAVEWATCH III hindcast on the Draupner rogue wave event. We measure ensemble statistical moments as functions of time, finding that although the crossing angle influences the statistical evolution to some degree, there are no significant third order effects present. Furthermore, we inquire on the nature of mean sea level measured beneath extreme crest heights, the elevation of which (set up or set down) is shown to be related to the spectral content in the low wavenumber region of the corresponding spectrum. 

\section{Crossing sea model}

\begin{figure}
\centering
\includegraphics[width=0.95\textwidth]{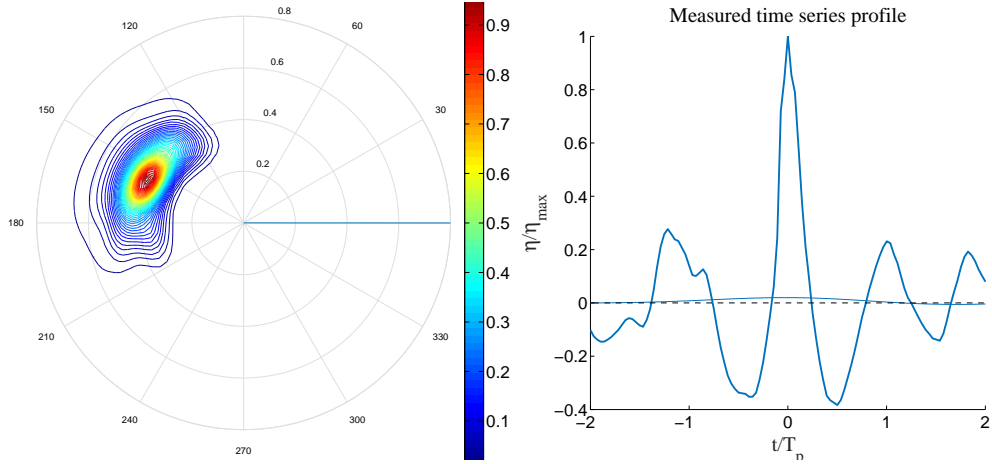}\\
\caption{Left panel: WAVEWATCH III hindcast directional wave spectrum $S(\omega,\theta)$ used as input for the HOS simulations (Draupner). Here, $\omega$ is the angular frequency and $\theta$ the direction in degrees. The spectrum has been normalized with respect to the spectral peak value. Right Panel: Measured Draupner time series. Thick line: Crest profile. Thin line: mean sea level. Dashed line: zero sea level. Note the positive mean sea level beneath the crest. Here, $\eta/\eta_{\max}$ is the surface displacement relative to its maximum value and $t$ is the time relative to the dominant wave period $T_p$ (which is 14.93 seconds for the Draupner spectrum).}
\label{fig:dsts} 
\end{figure}

To efficiently perform high resolution, phase resolved simulations of the evolution of crossing sea states, we employ the Higher Order Spectral (HOS) numerical method of West \textit{et al.} (1987), to solve the free surface Euler equation system with a flat sea bottom. This requires appropriate initial conditions for the free surface $\eta(x,y,t)$ and associated surface velocity potential $\psi(x,y,t)$, which is obtained as follows. We begin by utilising a hindcast directional spectrum $S(\omega, \theta)$ for the Draupner wave event (Fig. \ref{fig:dsts}), produced using WAVEWATCH III, which is the same hindcast spectrum used in Fedele \textit{et al.} (2016). The Draupner wave itself was measured by Statoil at the Draupner oil platform on the $1^{st}$ of January, 1995, where the water depth is $70\,$m. 

It is important to recognise that nonlinear wave-wave interactions are modelled in WAVEWATCH III by the Hasselmann equation, and thus non-resonant interactions are not included (see Tolman \textit{et al.} (2014)). As such, the coupling between the HOS method and WAVEWATCH III is advantageous in the sense that the latter can perform wave hindcast or forecast on a global scale, over large periods, which is not feasible due to the resources required to do so with the HOS method. The HOS method provides higher resolution simulations and a more complete picture of the nonlinear wave wave interactions. Note that we use a third order HOS expansion in our simulations, which is equivalent to the Zakharov Hamiltonian formalism. 

Given that this hindcast spectrum does not feature a crossing sea state, we use it to artificially construct such a spectrum from the base hindcast by summing together our hindcast Draupner spectrum, with an identical copy of itself rotated by a crossing angle $\Omega = 90^{o}$. Additionally, we consider two other crossing angles of $45^{o}$ and $22.5^{o}$, which allow us to see the influence of $\Omega$ on the sea state evolution. Note that we refer below to this case as ``Draupner-Draupner (DD)". To simulate spectral growth between wind sea and swell, we also construct a crossing sea state using again the Draupner spectrum (swell) and a narrowband JONSWAP spectrum (wind), for the same crossing angles as before. The JONSWAP peak is constructed using the following parameters: peak shape parameter $\gamma = 10$, angular frequency, $\omega_p = 0.8 \omega_{p_D}$, $H_s = 1.1 H_{s_D}$, where the subscript $D$ indicates Draupner hindcast values, and the subscript $p$ implies dominant (or peak) wave modes. The directional spread of the JONSWAP spectrum is set to $20^{o}$. Note that we refer below to this case as ``Draupner-JONSWAP (DJ)''.

Once we have obtained the directional spectrum $S(\omega, \theta)$, we convert to a cartesian wavenumber coordinate system $(k_x, k_y)$ via the Jacobian transform
\begin{equation}
S(k_x,k_y) = \frac{1}{k}\frac{d\omega(k)}{dk}S(\omega(k),\theta), 
\end{equation}
where $k = |\mathbf{k}|$, and the direction is assumed to be $\theta = \tan^{-1}(k_y/k_x)$. The initial wavenumber spectra for each case are presented in Fig. \ref{fig:init_spec}.
 \begin{figure}
\centering
\caption{Initial crossing spectra used in the present HOS simulations, plotted in cartesian coordinates (as defined in the text). Results are shown for the Draupner-Draupner and Draupner-JONSWAP cases for different crossing angles.} 
\label{fig:init_spec} 
\end{figure}
 
Random phase is introduced via the random variable $\beta$, uniformly distributed over $[0,2\pi]$, and finally, the Fourier spectra for $\eta$ and $\psi$ are given by

\begin{eqnarray}
\hat{\eta}(k_x,k_y) =& \; \exp(i\beta)\sqrt{S(k_x,k_y)} + c.c., \\
\hat{\psi}(k_x,k_y) =& \; -i\sqrt{\frac{g}{k\tanh(kd)}}\exp(i\beta)\sqrt{S(k_x,k_y)} + c.c.,
\end{eqnarray}
where we have used linear theory to construct $\hat{\psi}$. The physical initial conditions are then recovered via an inverse Fourier transform. Note that this initial condition is linear, and so to ensure stable initial evolution, the nonlinear terms in the evolution equations are smoothly introduced via the Dommermuth ramping function (Dommermuth 2000), over a period $T_r \approx O(5T_p)$, where $T_p$ is the peak wave period of the spectrum. Furthermore, we implement the phenomenological based filter proposed by Xiao \textit{et al.} (2013) to account for energy dissipation due to wave breaking:
\begin{equation}
F(\mathbf{k}|k_p, f_1, f_2) = \exp\left(-\bigg{\vert} \frac{\mathbf{k}}{f_1k_p} \bigg{\vert}^{f_2}  \right),
\end{equation}
with parameters set as $[f_1, f_2] = [8,30]$.

Simulations are performed using $1024 \times 1024$ dealiased Fourier modes. The wave fields are scaled as $L_{x,y} = O(\epsilon^{-2} k_p^{-1})$, based on the Benjamin--Feir scale, and simulations run for times similarly chosen as $T = O(\epsilon^{-2} T_p)$. 

\section{Numerical simulation results}

Ensemble statistics are measured for each crossing sea spectrum considered.  The temporal evolution of excess kurtosis and skewness are computed from an ensemble of 20 simulations, for each case. Note that each member of the various ensembles is initialised with a newly generated random phase (i.e,. we are performing a Monte Carlo type anaylsis). Skewness is correlated with three wave interactions, while kurtosis is correlated to four wave interactions, in particular the Benjamin--Feir instability. There is a known relationship between kurtosis and the Benjamin--Feir index (BFI), a measure of a wave systems susceptibility to the instability. Thus, knowledge of these statistical moments can be used to infer the strength of second and third order nonlinearities. Onorato \textit{et al.} (2006) and Toffoli \textit{et al.} (2011) performed similar simulations using narrowband JONSWAP spectra, measuring the evolution of kurtosis as a function of crossing angle $\Omega$. By increasing $\Omega$, they found increased values of kurtosis, with maximum kurtosis occurring for $40^{o} < \Omega < 60^{o}$, in agreement with theoretical work on the CNLS system.

\begin{figure}
\centering
\includegraphics[width=0.95\textwidth]{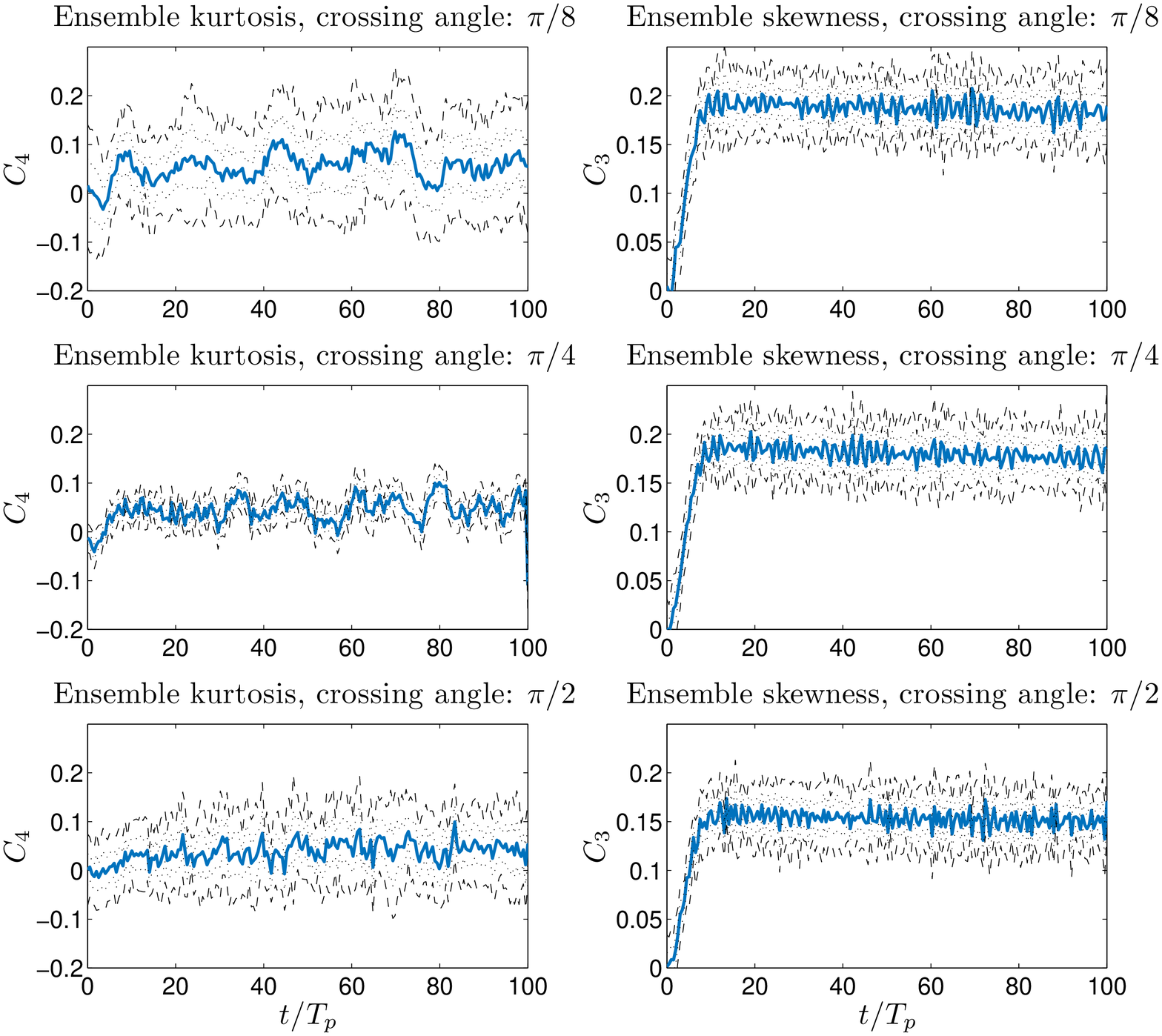}\\
\caption{Ensemble (averaged over 20 simulations each) evolution of kurtosis, $C_4$, and skewness, $C_3$, for Draupner-Draupner simulations. Solid blue lines: ensemble averaged measurements, dashed black lines: ensemble variance, dotted black lines: $95\%$ confidence intervals.}
\label{fig:dd_stat} 
\end{figure}

\begin{figure}
\centering
\includegraphics[width=0.95\textwidth]{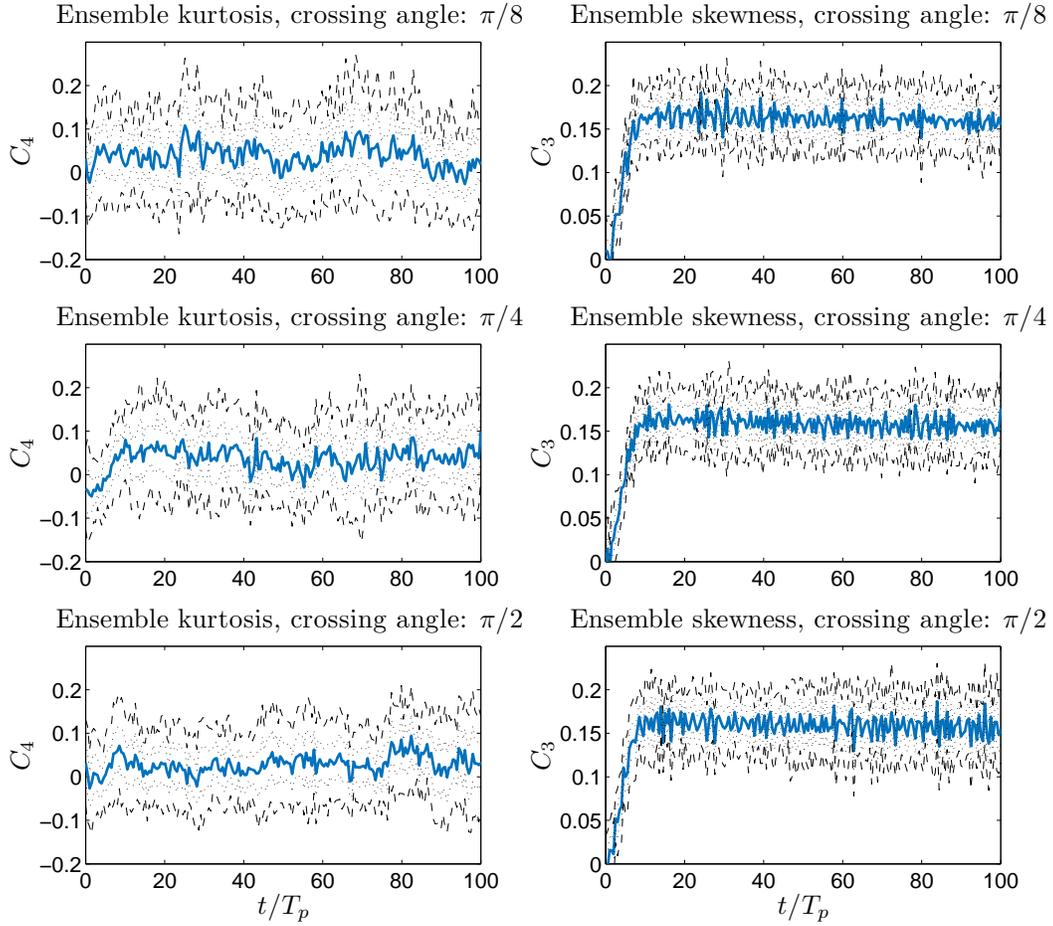}\\
\caption{Ensemble (averaged over 20 simulations each) evolution of kurtosis, $C_4$, and skewness, $C_3$, for Draupner-JONSWAP simulations. Solid blue lines: ensemble averaged measurements, dashed black lines: ensemble variance, dotted black lines: $95\%$ confidence intervals.}
\label{fig:dj_stat} 
\end{figure}

Ensemble statistical measurements for all crossing cases are analogous with the measurements from the singular Draupner hindcast simulations of Fedele \textit{et al.} (2016). The ensemble averages of kurtosis and skewness are shown in Fig. \ref{fig:dd_stat} (Draupner-Draupner case) and Fig. \ref{fig:dj_stat} (Draupner-JONSWAP case), while the mean measurements are given in Table \ref{tb:cross_stats}. These statistical measurements are measured spatially from each snapshot from the temporal evolution of the wave field. It is clear from the table that decreasing the crossing angle towards $22.5^{\circ}$ actually leads to increased kurtosis and skewness, contrasting with the narrowband simulation results mentioned above, though this is not surprising, given that the particular phenomenon is associated with the Benjamin--Feir instability. This effect is more dramatic for the Draupner-Draupner cases. The presence of a crossing spectral peak does not seem to influence significantly the nature of the statistical moments, at least for these broad banded oceanic spectra. 

\begin{table}
  \begin{center}
\def~{\hphantom{0}}
  \begin{tabular}{lccc}
      \textbf{Crossing Spectrum} & \textbf{Mean kurtosis}   &   \textbf{Mean skewness} \\[3pt]
    DD: $\Omega = 22.5^{\circ}$ & 0.0588 & 0.1863 \\ 
    DD: $\Omega = 45^{\circ}$    & 0.0462 & 0.1792 \\ 
    DD: $\Omega = 90^{\circ}$    & 0.0395 & 0.1520 \\ 
    DJ: $\;\Omega = 22.5^{\circ}$  & 0.0384 & 0.1612 \\ 
    DJ: $\; \Omega = 45^{\circ}$     & 0.0386 & 0.1577 \\ 
    DJ: $\; \Omega = 90^{\circ}$    & 0.0296 & 0.1582 \\ 
      \end{tabular}
 \caption{Mean values for kurtosis and skewness for Draupner-Draupner (DD) and Draupner-JONSWAP (DJ) crossing seas.}
\label{tb:cross_stats}
  \end{center}
\end{table}

\begin{figure}
\centering
\includegraphics[width=1\textwidth]{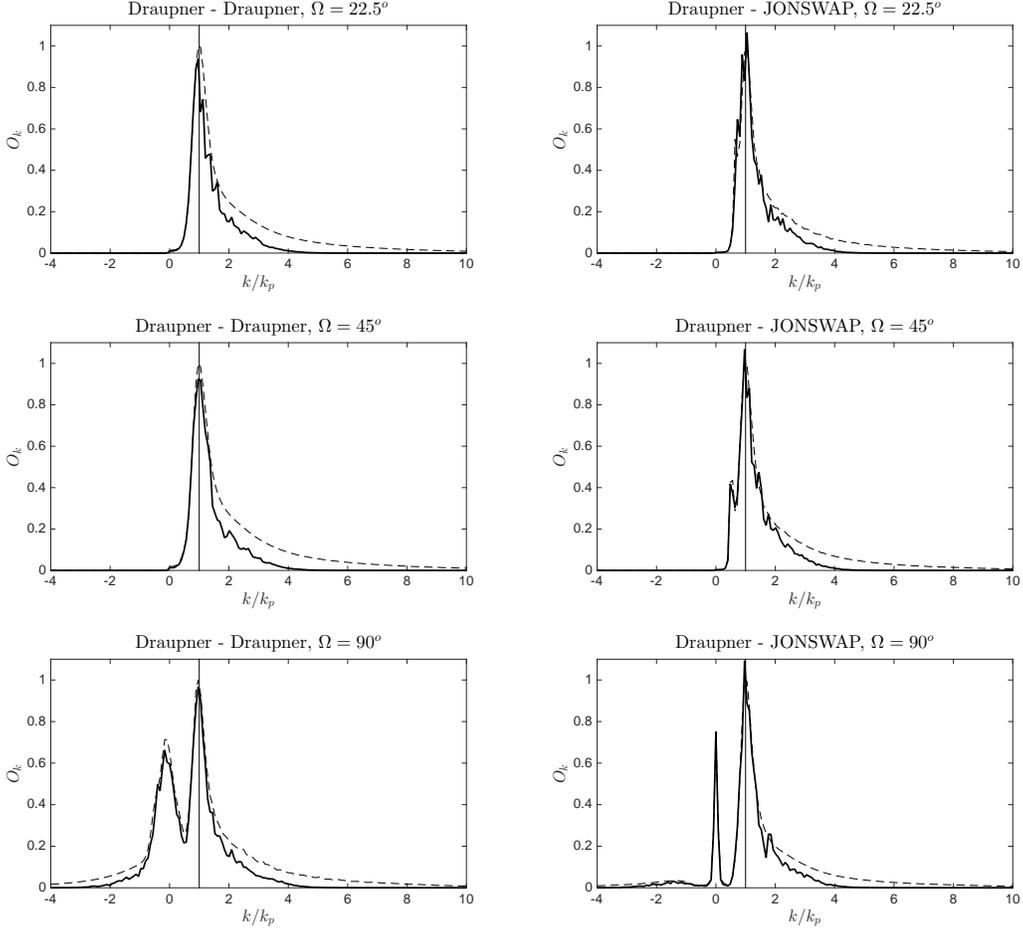}\\
\caption{Omnidirectional spectrum at the end of the simulations (solid lines), compared with the spectrum used as initial condition (dashed lines).} 
\label{fig:SK_plot} 
\end{figure}

\begin{figure}
\centering
\includegraphics[width=1\textwidth]{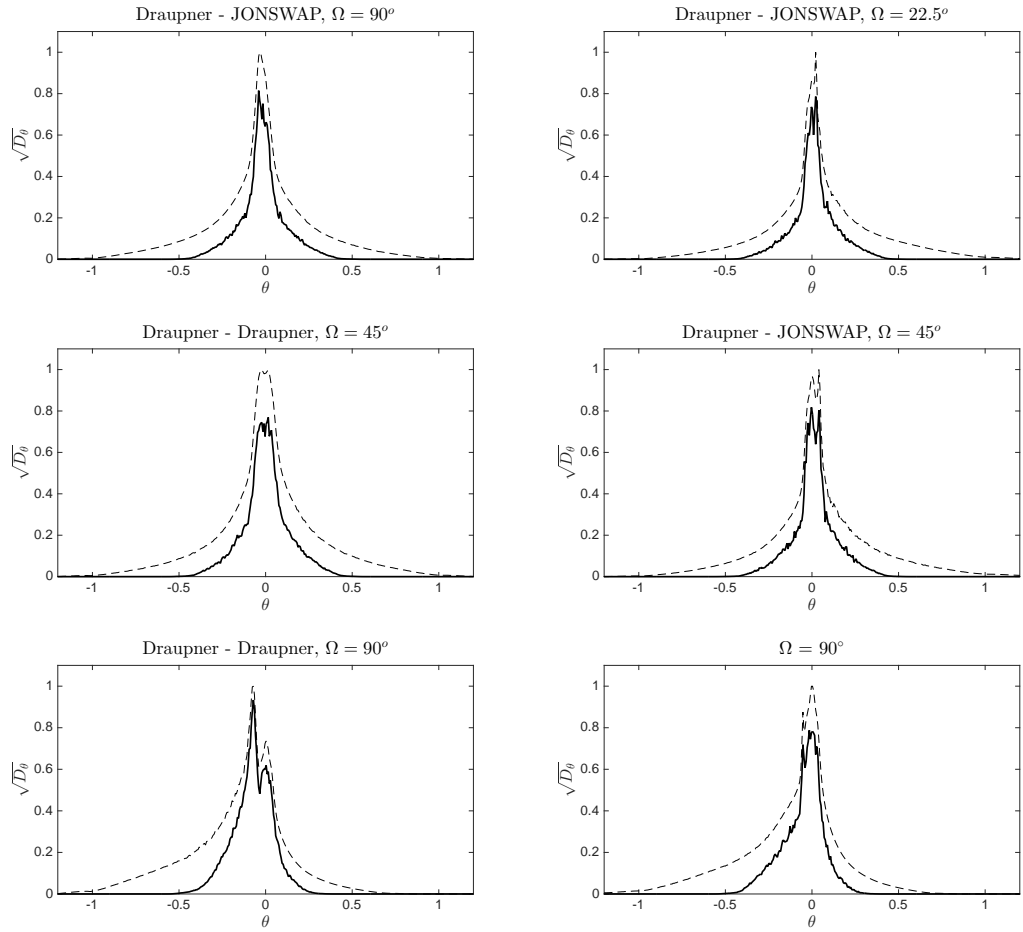}\\
\caption{Directional distribution at the end of the simulation (solid lines), compared with the initial condition (dashed lines). Note that we take the square root of the directional distribution, to emphasise the difference between initial condition and end of simulation.} 
\label{fig:DT_plot} 
\end{figure}

Next we inspect spectral growth, examining both the omnidirectional spectra $O_k(k)$(Fig. \ref{fig:SK_plot}) and directional distributions $D(\theta)$ (Fig. \ref{fig:DT_plot}) associated with each crossing case. We define them as follows: 
\begin{align}
O_k(k) = & \int_{\theta}kS(k,\theta)d\theta, & \\
D(\theta) = & \int_{0}^{\infty}S(k,\theta)dk. &
\end{align}
There is modest spectral distortion through the simulation, with some energy leaking from spectral sidebands to the peak. The Draupner-Draupner crossing case with $\Omega = 22.5^{\circ}$ is quite similar to the original Draupner simulation; given the small crossing angle and identical peak wavenumbers, this is not unexpected. Finally, there does not seem to be any clear case of one peak growing significantly at the expense of the other. In other words, there is little energy transfer between the peaks. Note that we normalise each spectrum by the initial peak value.

\begin{figure}
\centering
\includegraphics[width=1\textwidth]{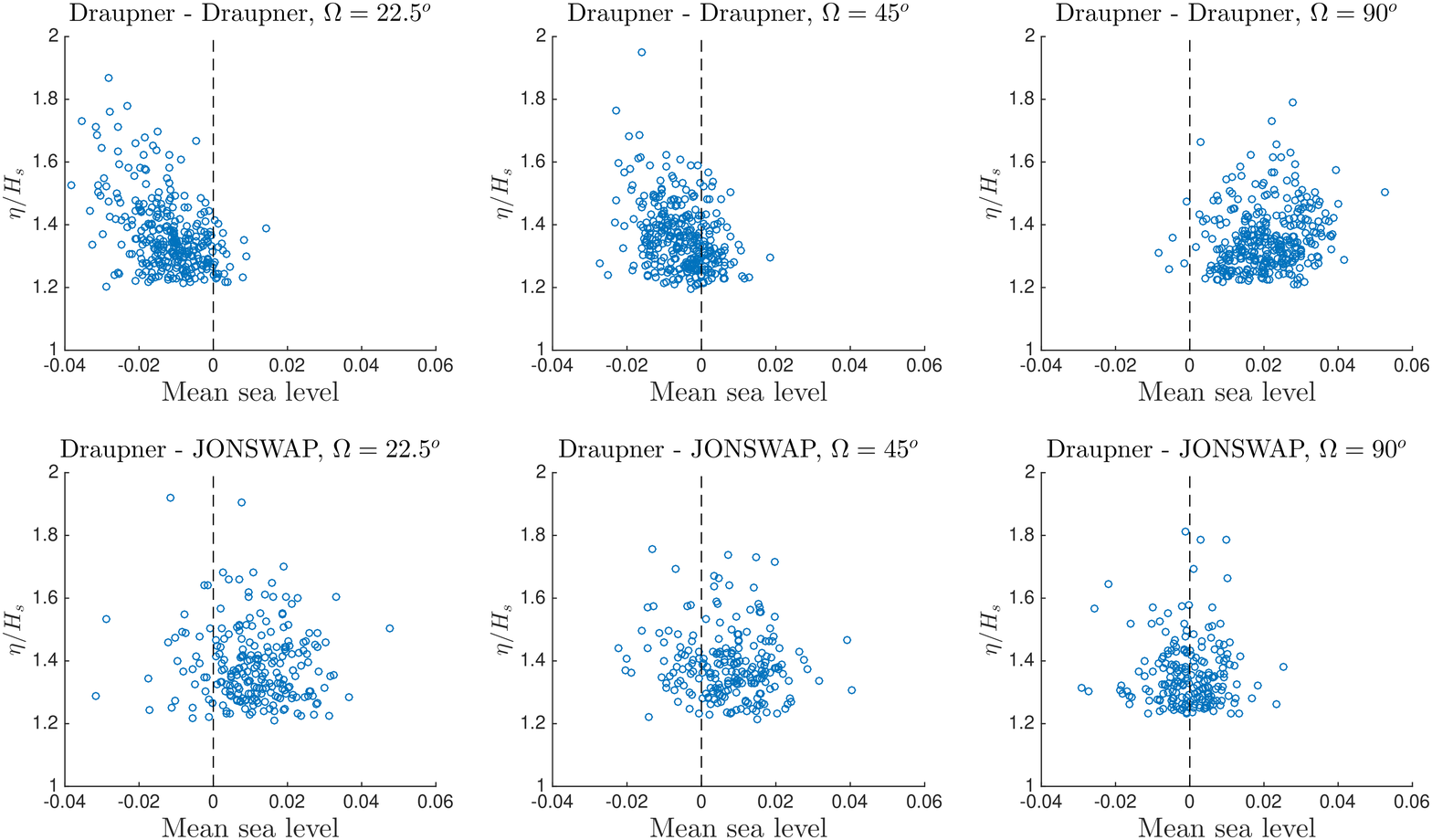}\\
\caption{Mean sea level versus crest height. Results are shown for the Draupner-Draupner and Draupner-JONSWAP cases for different crossing angles.} 
\label{fig:msl_plot} 
\end{figure}

Finally, for the various rogue waves recorded in our simulations (typically 10-15 waves recorded per ensemble member), we compare their measured crest height with the associated mean sea level (Fig. \ref{fig:msl_plot}). Rogue waves themselves are identified by measuring the local space-time maxima within each simulation, and taking those whose crest height exceeds $1.25$ times the significant wave height of the wave field at the time of the observation. As in Fedele \textit{et al.} (2016), mean sea level is estimated by low pass filtering the time series of each rogue crest below cut off frequency $f \sim f_p/2$. Interestingly, the crossing angle markedly influences the mean sea level beneath the measured extreme crests. For the Draupner-Draupner cases, increasing the angle leads to increased development of set up beneath the crest. In fact, for a crossing angle $\Omega = 90^{\circ}$, nearly all the measured crests coincided with a set up of mean sea level! As already mentioned, Adcock \textit{et al.} (2011) pointed out that the presence of a set up beneath the Draupner wave was possibly due to the perpendicular crossing of the spectrum with one of similar design. Remarkably, this phenomenon is reversed for the Draupner-JONSWAP cases. although we note that even for $\Omega = 22.5^{\circ}$ there is still a noteworthy portion of set up measurements. 

Comparing with Figs. \ref{fig:SK_plot} and \ref{fig:DT_plot}, it would appear that the prominence of a measured set up is related to the presence of a distinct second spectral peak in the low end of the wavenumber spectrum. Given that it is this part of the spectrum that will remain after the second order difference low pass filter is applied, an associated positive mean sea level is reasonable. The case which features this second peak most prominently is the $\Omega = 90^{\circ}$ Draupner-Draupner case, and moreover, all Draupner-JONSWAP cases contain this feature to some degree. For small crossing angles in the Draupner-Draupner regime, the peaks are simply not discernible enough. 

\section{Conclusion}
We have considered the evolution of crossing sea states, simulated via the higher order spectral method of West \textit{et al.} (1987). Based on speculation that the Draupner wave may have occurred in a crossing sea state, two different spectra are considered: a crossing between the original Draupner spectrum with itself at various crossing angles ($22.5^{\circ}, 45^{\circ},$ and $90^{\circ}$), and also, the Draupner spectrum with a JONSWAP `swell' system for the same angles. 

The statistical evolution of the studied wave fields is pointedly similar to the Draupner simulations found in Fedele \textit{et al.} (2016). Although it is known that crossing wave systems can possess enhanced growth rate of modulational instability, we note that if the isolated systems themselves are notably insusceptible to the instability (i.e., broad short crested ocean systems), the crossing of such systems likely will not stimulate it. Systems with small crossing angles seem to possess somewhat larger kurtosis and skewness, although the measurements are not outside the realms of possibility for wave fields with mono-peaked wave spectra. It would appear that, at least for the cases considered, that there is no extraordinary nonlinear third order interactions - at least those responsible for the Benjamin-Feir instability - when compared to regular oceanic sea states. We do note, however, that this does not rule out enhanced growth rates of the instabilities in the right circumstances, i.e., the crossing of narrow band spectra. As observed in Fedele \textit{et al.} (2016), evidence suggests that the evolution of crossing sea states in typical oceanic conditions is most likely dominated by second order nonlinearities, with extreme or rogue waves developing as a result of constructive interference enhanced by second-order bound nonlinearities.

By simulating these bi-peaked spectra, we have observed a possible explanation for the set up of mean sea level beneath the Draupner rogue wave. It would appear that prominent set up is connected to significant secondary peak in the low wavenumber portion of the wave field omnidirectional spectrum. Although we have induced this secondary peak by simulating crossing sea states, it is not implausible for it to develop naturally in a mono-peaked system. Thus, it transpires that a set up of mean sea level beneath a large wave crest is also not a remarkable feature, and may just be a consequence of the low wavenumber (frequency) portion of spectrum containing a relatively large proportion of the energy of the system. 
\clearpage
\section{Acknowledgements} 
This work is supported by the European Research Council (ERC) under the research projects ERC-2011-AdG
290562-MULTIWAVE and ERC-2013-PoC 632198-WAVEMEASUREMENT, and Science Foundation Ireland
under grant number SFI/12/ERC/E2227.

\end{document}